\documentclass[10pt]{article}
\usepackage{overcite,graphicx}

\makeatletter \def\@cite#1{\mbox{$\m@th^{\hbox{\@ove@rcfont[#1]}}$}}

\def\eqref#1{(\ref{#1})}

\def\thefigures#1{\par\clearpage\section*{Figures}\list
  {FIG.~\arabic{enumi}.}{\labelwidth\parindent\advance\labelwidth -\labelsep
      \leftmargin\parindent\usecounter{enumi}}}
\def\figitem#1{\item\label{#1}}

\newenvironment{eq}[1]%
{\begin{bequation}{#1}}{\end{bequation}}

\newenvironment{eqarray}[1]%
{\begin{beqnarray}{#1}}{\end{beqnarray}}

\newcommand{\inverse}[1]{ {1  \over {#1}} }

\newcommand{\twid}{\sim}

\newcommand{\paren}[1]{\left( #1 \right)}
\newcommand{\square}[1]{\left[ #1 \right]}
\newcommand{\curly}[1]{ \left\{ #1 \right\} }

\newcommand{\casesbracketsii}[4]
{\left\{
\begin{array}{ll}
#1\   & (#2) \\ & \blank #3\  & (#4) 
\end{array}%
\right.
}

\newcommand{\gap}{\hspace{.4in}}

\newcommand{\ggap}{\hspace{.8in}}

\newcommand{\blank}{\ \\}

\newcommand{\gt}{\rightarrow}

\newcommand{\period}{\ \ .}
\newcommand{\comma}{\ ,\ }

\newcommand{\const}{\mbox{const.}\ }

\newcommand{\ignore}[1]{}

{\begin{tabular}{#1}}%
{\end{tabular}}

\newcommand{\vi}{\vspace{.25in}}
\newcommand{\vii}{\vspace{.5in}}
\newcommand{\viii}{\vspace{.75in}}

{\end{list}}

{\end{list}}

{\ \\ XXXXX \hfill XXXXX #1 XXXXX\begin{equation}\label{#1}}%
{\end{equation}}

{\ \\ XXXXX \hfill XXXXX #1 XXXXX\begin{eqnarray}\label{#1}}%
{\end{eqnarray}}

\newenvironment{bequation}[1]%
{\begin{equation}\label{#1}}%
{\end{equation}}

\newenvironment{beqnarray}[1]%
{\begin{eqnarray}\label{#1}}%
{\end{eqnarray}}

\newcommand{\drop}{\nonumber \\}

\newcommand{\ddrop}{\drop\drop}

\newcommand{\ie}{i.\,e.\ }

\newcommand{\phidot}{\dot{\phi}}
\newcommand{\phibroad}{\phi_{\small broad}}
\newcommand{\phibroaddot}{\dot{\phi}_{\small broad}}
\newcommand{\Psidot}{\dot{\Psi}}
\newcommand{\Psicoh}{\Psi_{\small coh}}
\newcommand{\Psiinc}{\Psi_{\small inc}}
\newcommand{\phidead}{\phi_{\small dead}}
\newcommand{\phideaddot}{\dot{\phi}_{\small dead}}
\newcommand{\phideaddotbar}{\overline{\dot{\phi}}_{\small dead}}
\newcommand{\phideadideal}{\phi_{\small dead}^{\small ideal}}

\newcommand{\phideadidealdotbar}{\overline{\dot{\phi}}_{\small dead}^{\small ideal}}

\newcommand{\phideadtransdotbar}{\overline{\dot{\phi}}_{\small dead}^{\small trans}}

\newcommand{\phideadcombcrossdotbar}{\overline{\dot{\phi}}_{\small dead,cross}^{\small comb}}
\newcommand{\phideadbroad}{\phi_{\small dead}^{\small broad}}

\newcommand{\phideadcombdotbar}{\overline{\dot{\phi}}_{\small dead}^{\small comb}}
\newcommand{\phideaddispdotbar}{\overline{\dot{\phi}}_{\small dead}^{\small disp}}
\newcommand{\phideadread}{\phi_{\small dead}^{\small read}}
\newcommand{\phicoh}{\phi_{\small coh}}
\newcommand{\phiinc}{\phi_{\small inc}}
\newcommand{\phicohdot}{\dot{\phi}_{\small coh}}
\newcommand{\phiincdot}{\dot{\phi}_{\small inc}}
\newcommand{\phideadinc}{\phi_{\small dead}^{\small inc}}
\newcommand{\phideadincdotbar}{\overline{\dot{\phi}}_{\small dead}^{\small inc}}
\newcommand{\phienv}{\phi_{\small env}}
\newcommand{\vp}{v_{\small p}}
\newcommand{\kp}{k_{\small p}}
\newcommand{\kc}{k_{\small c}}
\newcommand{\kd}{k_{\small d}}
\newcommand{\ctr}{c_{\small tr}}
\newcommand{\ktr}{k_{\small tr}}

\newcommand{\Ri}{R_i}
\newcommand{\Ribar}{\bar{R}_i}
\newcommand{\Ritilde}{\widetilde{R}_i}
\newcommand{\Htilde}{\widetilde{H}}
\newcommand{\Nbar}{\bar{N}}
\newcommand{\Deltatilde}{\widetilde{\Delta}}

\newcounter{fignumber}
\begin{document}


\renewcommand{\thepage}{}

\begin{center}

\LARGE 
{\bf Pulsed Laser Polymerization at \\  
Low Conversions: \\
Broadening and Chain Transfer Effects}

\viii

\Large
BEN O'SHAUGHNESSY$^*$ 
and
DIMITRIOS VAVYLONIS \\
Department of Chemical Engineering\\
Columbia University \\ 
500 West 120th Street \\
New York, NY 10027, USA

\end{center}

\normalsize


\ \\ \newline
This is a preprint of an article accepted for publication in \\
Macromolecular Theory and Simulations \\ \copyright 2003 Wiley-VCH Verlag
GmbH
\ \\ \ \\ \ \\
$^*$\ To whom correspondence should be addressed.\\  

\pagebreak


\pagenumbering{arabic}

\section*{SUMMARY}

Pulsed laser polymerization (PLP) is widely employed to measure
propagation rate coefficients $\kp$ in free radical polymerization.
Various properties of PLP have been established in previous works,
mainly using numerical methods. The objective of this paper is to
obtain analytical results.  We obtain the most general analytical
solution for the dead chain molecular weight distribution (MWD) under
low conversion conditions which has been hitherto obtained.
Simultaneous disproportionation and combination termination processes
are treated.  The hallmarks of PLP are the dead MWD discontinuities
located at integer multiples of $n_0 = \kp t_0 C_{\rm M}$, where $t_0$
is the laser period and $C_{\rm M}$ is the monomer concentration.  We
show that chain transfer reduces their amplitude by factors $e^{-\ctr
L n_0}$, consistent with numerical results obtained by other workers.
Here $\ctr$ is the chain transfer coefficient and $L n_0$
($L=$integer) are the discontinuity locations.  Additionally, transfer
generates a small amplitude continuous contribution to the MWD.  These
results generalize earlier analytical results which were obtained for
the case of disproportionation only.  We also considered 2 classes of
broadening: (i) Poisson broadening of growing living chains and (ii)
intrinsic broadening by the MWD measuring equipment (typically gel
permeation chromatography, GPC).  Broadening smoothes the MWD
discontinuities.  Under typical PLP experimental conditions, the
associated inflection points are very close to the discontinuities of
the unbroadened MWD.  Previous numerical works have indicated that the
optimal procedure is to use the inflection point to infer $\kp$.  We
prove that this is a correct procedure provided the GPC resolution
$\sigma$ is better than $n_0^{1/2}$. Otherwise this underestimates $L
n_0$ by an amount of order $\sigma^2/n_0$.

\vii

{\bf Keywords}: molecular weight distribution/propagation rate
coefficients/pulsed laser polymerization/radical polymerization/theory

\pagebreak


\section{Introduction}

A fundamental material parameter in free radical polymerization (FRP)
is the propagation rate constant
\cite{flory:book,odian:book,iupac:frp_3} $\kp$ governing the
sequential addition of monomers to the active free-radical ends of
``living'' chains (see fig. \ref{propagation}).  In this paper we
consider theoretically ``pulsed laser polymerization'' (PLP)
\cite{beuermannbuback:plp_review_2002,coote:plp_review}, possibly
the most accurate method to to measure $\kp$.  In PLP new radicals
(living chains of one monomer, $N=1$) are generated by a sequence of
laser pulses which photocleave photoinitiators in the reaction mix.
The laser is flashed periodically (the duration of each flash is
extremely short (see fig. \ref{ideal_plp}(a))) with flash period $t_0
\twid 0.1$sec optimally chosen to be somewhat shorter than the average
living chain lifetime.  Thus every $t_0$ the living chain population
is updated by the injection of new primary radicals.  After many
cycles a periodic state is established where the time averaged rate of
radical production balances the time averaged rate of living chain
termination: pairs of living chains terminate to generate either a
single (``combination'') or a pair (``disproportionation'') of
``dead'' chains, the final polymer product (see fig.
\ref{termination}).

Now unlike the dead molecular weight distribution (MWD) of standard
steady state FRP which has no singular features, in the case of PLP
the dead MWD, $\phidead(N)$, is predicted theoretically to possess
finite discontinuities \cite{iupac:frp_3,olaj:plp} at chain lengths
$N = L n_0$, where $L=0,1,2,...$ and
                                                \begin{eq}{cook}
n_0 \equiv \vp t_0 \comma \gap
\vp \equiv \kp C_{\rm M}
\period
                                                                \end{eq}
Here $n_0$ is the number of monomers added to a growing living chain
in time $t_0$ and the ``propagation velocity'' $\vp$, namely the rate
of monomer addition to a growing chain, is proportional to the monomer
concentration $C_{\rm M}$.  The origin of the dead MWD discontinuities
is the coherent sudden increase in termination rates due to the sudden
increase in radical population following each pulse.  In principle the
discontinuities can be seen experimentally after measuring the
resulting MWD by gel permeation chromatography (GPC).  Thus $n_0$ and
hence $\vp$ (or equivalently $\kp$) is inferred.  This method has been
employed to deduce $\kp$ for a variety of polymerizing systems
\cite{hutchinson:plp:methacrylates,hutchinson:plp:methacrylates2,%
hutchinson:plp:alkylmethacrylates,hutchinson:plp,%
hutchinson:plp:vinyl_acetate,shipp:plp:methacrylonitrile}.

In reality there exist a number of practical difficulties.  (1)
Broadening effects.  Consider a living chain created at $t=0$.  After
time $t_0$ its length is not exactly $\vp t_0$; since polymerization
is a random process, Poisson fluctuations arise around this length.
This then leads to a softening of the theoretically infinitely sharp
dead MWD discontinuities.  Another, but quite different effect, is the
direct broadening of the dead MWD due to intrinsic uncertainties in
the measuring method.  GPC equipment if supplied with a discontinuous
dead MWD will output a reading indicative of a somewhat broadened
discontinuity due to fluctuations in the transit time of a chain of a
given length through the gel.  (2) Chain transfer \cite{flory:book}
of radicals from living chain ends to surrounding unpolymerized
monomer (fig. \ref{transfer}), quantified by the parameter $\lambda$.
This is simply related to the chain transfer coefficient $\ctr$:
                                                \begin{eq}{c}
\ctr \equiv {\ktr C_{\rm X} \over \kp C_{\rm M}} = {\lambda \over \vp}
\comma
                                                                \end{eq}
where $\ktr$ is the 2nd order rate constant governing chain transfer
to species X whose concentration is $C_{\rm X}$ (X is usually the
monomer species).  One expects chain transfer, important for species
such as vinyl acetate and polystyrene
\cite{hutchinson:plp:vinyl_acetate,hutchinson:plp:chaintransfer}, 
to dephase the living MWD which would otherwise be a sequence of
in-phase pulses.  This will then affect the discontinuities in the
dead MWD.

Typical experimental MWDs produced by PLP are shown in fig.
\ref{hutchmwd} where an obviously broadened first discontinuity is
evident near molecular weight 40,000.  In order to deduce the correct
propagation rate constant, it is essential to have an algorithm to
infer $n_0$ from such a broadened discontinuity.  Taking $n_0$ as the
molecular weight corresponding to the lower point of inflection, near
the middle of the broadened discontinuity, has been proposed as the
optimal approach \cite{olaj:plp,ziffererolaj:plp:thermalinit}.
Other authors have advocated that under particular conditions this
criterion should be replaced with the maximum point
\cite{sarneckischweer:plp,buschwahl:plp_transfer}.

In order to design and interpret PLP experiments many researchers have
employed numerical simulations of PLP chemical kinetics.  By varying
the parameters of the simulations the resulting dead MWD is modeled
under different experimental conditions.  In such numerical studies
Poisson-broadening is automatically taken into account and numerous
side-reactions such as chain transfer can be incorporated.  Guidelines
can then be derived for extracting $\kp$ from the broadened MWD peak.
Such numerical approaches have so far been employed both in the
absence
\cite{hutchinson:plp:vinyl_acetate,buback:plp_simul,odriscollkuindersma:plp_montecarlo}
and the presence of chain
transfer\cite{hutchinson:plp:chaintransfer,yan:plp_ch_transfer_numerical,buschwahl:plp_transfer}.
The effects of experimental broadening have been addressed numerically
in refs.
\cite{buback:plp_simul,buschwahl:plp_transfer}.

Analytical solutions of PLP dynamics are also very desirable. Where
available they provide a systematic and quantitative framework which
exactly articulates the conclusions of the underlying physical model,
beyond the scope of numerical analyses.  Analytical expressions can
also greatly facilitate experimental design by obviating the need for
numerical work.  The first analytical solution of PLP dynamics
appeared in refs.  \cite{olaj:plp,alexandrov:plp} and included the
effects of Poisson broadening for termination by both combination and
disproportionation.  A new derivation and a review of these
theoretical results is given in
ref. \cite{kornherr:olaj_plp_revisited}.  A closed form for the
dead MWD for arbitrary sequences of laser pulses was derived in
ref. \cite{evseevnikitin:plp_arbitrary_sequence}.  Analytical work
including chain transfer exists for the case of termination by
disproportionation only
\cite{ziffererolaj:plp:chaintrans1,ziffererolaj:plp:chaintrans2}.

In this paper we develop a completely general analytical theory for
PLP, including the effects of broadening and chain transfer
analytically.  We thus quantify the effect of these two mechanisms on
the theoretically infinitely sharp dead MWD which enables us to
discuss the validity of empirical rules on deriving $\kp$ from the
broadened MWD peak.  We consider a general (and typical) situation
where both combination and disproportionation events occur.  This
extends previous analytical studies.  We derive the most general
analytical expressions which have been obtained hitherto for PLP, to
the best of our knowledge.  Closed expressions are derived in terms of
the following independent parameters: $\vp$, $\kc$, $\kd$, $\lambda$,
$t_0$, and the concentration of radicals produced per laser pulse.
Here $\kc$ and $\kd$ are the combination and disproportionation rate
constants, respectively (see fig. \ref{termination}).  An important
quantity in what follows is the {\em net} termination rate constant,
                                                \begin{eq}{dog}
k \equiv \kc + \kd \period
                                                                \end{eq}

Our analysis describes low conversion FRP which allows us to make the
approximation that the net living-living bimolecular termination rate
constant, $k$, is independent of chain length (see fig.
\ref{termination}).  In reality $k(M,N)$ depends on the degrees of
polymerization $M,N$ of the reacting chains.  However in dilute
solutions (under good solvent conditions) theory
\cite{ben:interdil_letter,ben:interdil,ben:review_ijmp} predicts a
very weak dependence on chain lengths: $k(N,N) \twid N^{-\alpha}$ with
$\alpha \approx 0.16$.  A very weak dependence of $k$ on chain lengths
has also been well established experimentally in phosphorescence
quenching studies \cite{odriscollmahabadi:sip}.  At low
conversions, living chain terminations occur in a pure monomer solvent
(usually a good one) and thus the approximation of constant $k$ is
expected to yield rather accurate conclusions.  Making this
approximation, all quantities can be calculated exactly.  The effect
of the dependence of termination rates on chain length has been
addressed numerically in refs.
\cite{olaj:plp:kt2,olaj:plp:kt3,olaj:poisson_plus_kt,olaj:kt4,%
nikitinevseev:kt,nikitinevseev:kt_kp,odriscollkuindersma:plp_montecarlo}
which employed various empirical rules for the form of $k(M,N)$.

The living MWD, $\phi(N,t)$, like all quantities settles down to a
periodic form with period $t_0$.  Bimolecular reactions as in figs.
\ref{termination}, \ref{transfer} generate dead chains of length $N$ at a rate
                                                \begin{eq}{parrot}
\phideaddot(N,t) = {1 \over 2} \kc \int_0^N dM \phi(M,t) \phi(N-M,t)
\ +\  \kd \phi(N,t) \, \Psi(t)\  + \  \lambda \, \phi(N,t) 
\comma
                                                                \end{eq}
where $\Psi$ is the total concentration of living chains.  The
resulting dead MWD will include Poisson broadening effects provided
one uses the correct Poisson broadened living MWD in eq.
\eqref{parrot}.  In the following we will employ the notation $\phidead$ to
denote the dead MWD without Poisson broadening, including however
chain transfer effects (\ie $\phidead$ results from using in eq.
\eqref{parrot} the form for $\phi(N,t)$ which correctly includes chain
transfer but which neglects Poisson broadening).  The full Poisson
broadened MWD is denoted $\phideadbroad$ while the reading output by
the GPC measuring equipment we name $\phideadread$.

The dead MWD is calculated in successive steps as follows.  In section
2 we review ``ideal'' PLP, \ie we ignore chain transfer (setting
$\lambda = 0$) and broadening effects.  The time averaged dead MWD,
$\phideadideal$, turns out to have finite discontinuities at $N=j n_0$
($j$ is a positive integer).  In section 3 we generalize the results
of section 2 to include chain transfer and express the resulting dead
MWD, $\phidead$, in terms of $\phideadideal$.  We find that chain
transfer diminishes the amplitude of the dead MWD discontinuities, but
does not broaden them.  Broadening effects due to both Poisson
broadening and due to limited experimental resolution are included in
section 4.  We show that the inflection point rule is correct for
inferring $n_0$ only when the width of the Poisson broadened
discontinuity is larger than the experimental resolution.  We conclude
with a summary of our results in section 5.

Throughout, $[N]$ will denote the nearest integer to $N/n_0$ which is
less than $N/n_0$, whilst $rN \equiv {\rm rem}[N,n_0]$ denotes the
remainder of $N$ modulo $n_0$.  Similar notation will apply to time
$t$, with $n_0$ replaced by $t_0$.  Thus
                                                \begin{eq}{remainder}
N \equiv [N] n_0 + rN 
\comma \gap
rN \equiv {\rm rem}[N,n_0] ; \gap
t \equiv [t] t_0 + rt
\comma \gap
rt \equiv {\rm rem}[t,t_0]
\period
                                                                \end{eq}


\section{Ideal PLP}

Let us first briefly review the ``ideal'' PLP problem ignoring the
effects of broadening and chain transfer, as analyzed by Olaj et al. 
\cite{olaj:plp,kornherr:olaj_plp_revisited}.  Termination may occur
either by combination or disproportionation reactions.  First we
calculate the living MWD which depends on the total termination rate
constant, $k$, only, and not on the mode of termination.  We then
derive separate expressions for the dead chains terminated by
combination and disproportionation, respectively.

\subsection{Total Concentration of Living Chains $\Psi(t)$ and
Reaction Field $H(t)$}

The number density MWD of living chains, $\phi(N,t)$, obeys
                                                \begin{eq}{phi}
\phidot = - \vp \phi' - H(t) \phi + \delta(N) \Ri(t)
\comma
                                                                \end{eq}
with $\phi(N<0,t) = 0$.  Here $\phidot \equiv \partial \phi/\partial
t$ and $\phi' \equiv \partial \phi /\partial N$.  Note that the $\delta(N)
\Ri(t)$ term which injects primary radicals (chains of length $N=0$ in
our continuous framework) into the system at rate $\Ri(t)$ per unit volume, is
equivalent to the boundary condition $\vp \phi(0,t) = \Ri(t)$.  In PLP
dynamics the time-dependent primary radical production rate, $\Ri(t)$,
is a train of pulses,
                                                \begin{eq}{ri}
\Ri(t) = \Ribar t_0 \sum_{j=-\infty}^{\infty} \delta (t+j t_0)
\comma
                                                                \end{eq}
as shown in fig. \ref{ideal_plp}(a).  $\Ribar$ is the time averaged
radical production rate.  The $-\vp \phi'$ term in eq. \eqref{phi}
describes propagation whilst $H(t)\phi(N,t)$ is the total termination
rate (due to both combination and disproportionation) for a chain of
length $N$ due to the ``reaction field'' $H(t)$, namely the total
termination rate per unit time for a given chain due to all other
chains in the system:
                                                \begin{eq}{hpsi}
H(t) \equiv k \Psi(t) \comma \gap
\Psi(t) \equiv \int_0^\infty dN \phi(N,t)
\period
                                                                \end{eq}
Here $\Psi(t)$ is the total number density of living chains whose
dynamics follow after integrating eq. \eqref{phi} over all positive
$N$ values:
                                                \begin{eq}{psidynamics}
\Psidot = - H(t) \Psi + \Ri(t) = -k \Psi^2 + \Ri(t).
                                                                \end{eq}
For $t$ different from integer multiples of $t_0$ one has $\Ri(t)=0$
and thus $\Psi$ obeys simple second order reaction kinetics, $\Psidot
= - k\Psi^2$, with solution
                                                \begin{eq}{fat}
\Psi(t) = {\Psi_0 \over 1 + t k \Psi_0} \gap (0<t<t_0) \comma
                                                                \end{eq}
where $\Psi_0 \equiv \Psi(0^+)$ is the value of $\Psi$ just after a new
laser pulse at $t=0$.  We define $\beta$ as the surviving fraction
just before the next pulse at $t_0$,
                                                \begin{eq}{pick}
\beta \equiv \Psi(t_0^-) / \Psi_0
\period
                                                                \end{eq}
The values of $\Psi_0$ and $\beta$ when the system settles down to a
periodic stationary state are determined in terms of $\Ribar, k,$ and
$t_0$, by equating the number of terminated chains in one cycle with
the number of new living chains introduced by a pulse,
                                                \begin{eq}{wombat}
(1-\beta) \Psi_0 = \Ribar t_0 \period
                                                                \end{eq}
This relation together with eqs. \eqref{fat} and \eqref{pick} yields
                                                \begin{eq}{fact}
\Psi_0 = \inverse{2} \Ribar t_0 \square{1+ \{1+ 4/(\Ribar k t_0^2)\}^{1/2}}
\comma \gap
\beta = \inverse{1 + k \Psi_0 t_0}
\period
                                                                \end{eq}
The steady state $\Psi(t)$, sketched in fig. \ref{ideal_plp}(c), is of
course periodic with period $t_0$, and has discontinuous jumps 
$\Delta \Psi = (1-\beta) \Psi_0$ at integer multiples of $t_0$. 

\subsection{Living MWD}

The solution to eq. \eqref{phi} is derived in appendix A for a
general $\Ri(t)$ in terms of $H(t)$ whose important properties are
derived in appendix B.  Using the present periodic form for $\Ri(t)$,
eq. \eqref{ri}, using eqs. \eqref{wombat} and \eqref{hproperty}
and recalling the periodicity of $\Psi$ and $H$, we have
                                            \begin{eqarray}{living-pure}
\phi(N,t) 
 &=& {(1-\beta) \Psi_0} \Theta(N) \sum_{j=-\infty}^{\infty} 
e^{-\int_0^{N/\vp} H(t')dt'} \delta (N - \vp t - j n_0)
                                        \drop
&=& \sum_{j=-\infty}^{\infty} \phienv(N) \delta(N- \vp t - j n_0)
\comma
                                                                \end{eqarray}
where $\Theta$ is the step function ($\Theta(x>0)=1$,  $\Theta(x<0)
= 0$) and the function $\phienv$, sketched in fig. \ref{ideal_plp}(b), is
the living MWD envelope
                                                \begin{eq}{a}
\phienv(N) \ \equiv \ \Theta(N) \beta^{[N]} (1-\beta) \Psi(N/\vp)
\ = \ \Theta(N) \beta^{[N]} (1-\beta) {\Psi_0 \over 1 + k \Psi_0 rN / \vp}
\period
                                                                \end{eq}
The living MWD, sketched in fig. \ref{ideal_plp}(b), is a train of
pulses in phase with $\Ri(t)$.  A new $\delta$-pulse is born at $N=0$
at time $t=0$, say, and then moves in the +$N$ direction with velocity
$\vp$.  When in the interval $(j-1)n_0 < N \le j n_0$ (the ``$j$th
sector'') this pulse suffers the same field $H(t)$ as in all other
sectors.  The amplitude of the $\delta$-pulse in sector $j+1$ is thus
reduced by a factor $\beta$ with respect to sector $j$ and follows the
envelope $\phienv(N)$.

\subsection{Dead MWD}

The time-averaged rate of dead chain generation,
$\phideadidealdotbar(N)$, is the sum of the contributions due to
termination by combination and disproportionation,
$\phideadidealdotbar(N) = \phideadcombdotbar(N) +
\phideaddispdotbar(N)$.  Each of these two contributions can be found
by inserting eq. \eqref{living-pure} into the first and second terms on
the rhs of eq. \eqref{parrot}, respectively, and then time averaging
over one period.  In the following we will use however a simpler
method.

Consider first chains terminated by combination.  We notice that, due
to periodicity, the total concentration of dead chains of length $N$
produced by combination per period, $\phideadcombdotbar(N) t_0$,
equals one half the concentration of dead chains produced by
combination by one single $\delta$ pulse throughout its entire
lifetime.  Thus choosing the $j=0$ pulse in eq. \eqref{living-pure},
bimolecular kinetics imply
                                                \begin{eq}{statphys}
\phideadcombdotbar(N) = {\kc \over 2 t_0} \int_{-\infty}^{\infty} dt 
        \int_{-\infty}^{\infty} dM \, \phienv(M) \delta(M-\vp t)
\phi(N-M,t)
\period
                                                                \end{eq}
This is proved rigorously in Appendix C.  Since the integrand of
eq. \eqref{statphys} is zero for negative $t$ and $M$ values, we have
extended the lower limits of  the integrals from $0$ to $-\infty$ for
future convenience.

Substituting the expression for $\phi$ of eq. \eqref{living-pure} into
the above expression, it is shown in Appendix D this leads to 
                                                \begin{eq}{rat}
\phideadcombdotbar(N) = C {\kc \over k } \curly{
(L+1) \beta^L f^2\paren{rN \over 2} + L \beta^{L-1} 
f^2\paren{rN + n_0 \over 2} 
} \comma
                                                                \end{eq}
where
                                                \begin{eq}{f}
N \equiv L n_0 + rN \comma \gap
f(n) \equiv 
{\beta \over \beta+(1-\beta) n/n_0} \comma \gap
C \equiv {k \Psi_0^2 (1-\beta)^2 \over 4 \vp t_0}
\period
                                                                \end{eq}

Similarly, the time-averaged concentration of dead chains generated by
disproportionation can be calculated by noting that
$\phideaddispdotbar(N) t_0$ is equal to the concentration of dead
chains of length $N$ produced by disproportionation by one single
$\delta$-pulse throughout its lifetime.  Choosing $j=0$ for this pulse
in eq. \eqref{living-pure} one has
                                                \begin{eq}{serb}
\phideaddispdotbar(N) = {\kd \over t_0}
\int_{-\infty}^{\infty} dt \ \phienv(N) \ \delta(N-\vp t) \ \Psi(t)
\period
                                                                \end{eq}
Integrating the $\delta$ function in eq. \eqref{serb} and using eq.
\eqref{a} to express $\phienv$ in terms of $\Psi$ yields
                                                \begin{eq}{disp-ideal}
\phideaddispdotbar(N) = 4 C {\kd \over k} 
\curly{{\beta^L \over 1-\beta} f^2(rN)} \comma \gap
N \equiv L n_0 + rN \period
                                                                \end{eq}

The dead MWD, $\phideadideal = t \phideadidealdotbar =
t(\phideadcombdotbar+\phideaddispdotbar)$, is shown in fig.
\ref{ideal_plp}(d).  Both combination and disproportionation
contributions have discontinuities at the end of each sector $L$
(where $L=0$ labels the first sector, $0 \le N \le n_0$).
The discontinuity, $\Delta \phideadideal$, at the boundary of sectors $L-1$
and $L$ is 
                                                \begin{eq}{disc}
{\Delta \phideadideal(L n_0) \over \phideadideal(0)}
\equiv
{\phideadideal(L n_0^+) - \phideadideal(L n_0^-) \over \phideadideal(0)}
= \beta^L {2+4\rho \over 1 + 4 \rho/(1-\beta)}
\comma \gap
\rho \equiv \kd/\kc
\period
                                                                \end{eq}


\section{Chain Transfer}

In this section we will study how chain transfer modifies the results
of section 2.  Each transfer event results in one dead chain of
length $N$ and one living chain of length $N=0$ as depicted in fig. \ref{transfer}.  

\subsection{Living MWD}

The living dynamics now include additional loss through transfer from
chains of length $N$.  These radicals are reinjected as a source term
at $N=0$:
                                                \begin{eq}{bag}
\phidot = -\vp \phi' - [H(t)+\lambda] \phi +  [\Ri(t) + \lambda \Psi(t)] \delta(N)
\period
                                                                \end{eq}
The total reinjection rate is $\lambda \Psi$ where $\lambda$ is
defined in terms of the chain transfer coefficient in eq.
\eqref{c}.  

Integrating eq. \eqref{bag} over all positive $N$ values ($\phi$ is
zero for negative $N$) one immediately sees that the total number
density of living chains $\Psi(t)$ obeys the same dynamics as without
chain transfer, eq. \eqref{psidynamics}.  This must be true, of
course, since transfer events conserve the total number of living
chains and $k$ is independent of chain length in the present
approximation.  The crucial point is that $\Psi(t)$ and $H(t) = - k
\Psi(t)$ appearing in eq. \eqref{bag} are the {\em same} periodic
functions of time as without chain transfer, eq. \eqref{fat}.  

Since $H(t)$ is a known function of time, the dynamics of eq.
\eqref{bag} are effectively linear and we can use the superposition
principle to write $\phi = \phicoh + \phiinc$ where 
            \begin{eqarray}{bar}
\phicohdot   & = & - \vp \phicoh' - [H(t)+\lambda] \phicoh 
              + \Ri(t) \delta(N) 
                                               \ddrop
\phiincdot   & = & - \vp \phiinc' - [H(t)+\lambda] \phiinc
              + \lambda \Psi(t) \delta(N)
\period
            \end{eqarray}
Here $\phicoh$ is the coherent part, which as we will immediately
see is a sum of pulses in phase with radical production, whilst
$\phiinc$ is the incoherent part which is out of phase due to random
transfer events. 

Replacing $H \gt H+\lambda$ in the general solution, eq.
\eqref{phisoln}, to the living dynamics without chain transfer, one
sees $\phicoh$ is $e^{-\lambda N/\vp}$ times its no-transfer form of eq.
\eqref{living-pure}.  Similarly replacing $H \gt H+\lambda$ and $\Ri
\gt \lambda \Psi$, one obtains $\phiinc$ after using eq.
\eqref{hproperty}.  One finds:
                                                \begin{eqarray}{phicoh}
\phicoh(N,t) &=& e^{- \lambda N /\vp}
\sum_{j=-\infty}^{\infty} \phienv(N) \delta(N - \vp t - j n_0)
\comma \gap
\Psicoh(t) = \Psi(t) {(1-\beta) e^{- \lambda t} \over 1 - \beta
e^{-\lambda t_0}} \comma
                                \ddrop
\phiinc(N,t) &=& 
{\lambda e^{-\lambda N/\vp} \over \vp} \Psi(t) 
\beta^{[t] - [t-N/\vp]} \Theta(N)
\comma \ggap \ \ 
\Psiinc(t) = \Psi(t) - \Psicoh(t) \comma
                                                                \end{eqarray}
where $\Psicoh$, $\Psiinc$ are the corresponding total concentrations
of chains in each part of the MWD.  $\phiinc$ is a series of
propagating pulses each of width $n_0$, the amplitude of which
decreases with molecular weight and time.  Notice however that unlike
$\phicoh$, $\phiinc$ extends over all $N$ values at a given time.
$\phiinc$ has discontinuities at $N = \vp t + j n_0$.  At each
discontinuity, $\phiinc$ decreases suddenly by a factor $\beta$.

\subsection{Dead MWD}

Dead chains are generated by both living chain termination (first two
terms on the rhs of eq. \eqref{parrot}) and chain transfer events
(last term of eq. \eqref{parrot}).  We consider first the contribution
due to termination by combination.  Using the results of Appendix C,
the time averaged rate of production of dead chains by combination is
given by
                                                \begin{eq}{light}
\phideadcombdotbar(N) 
=
{\kc \over 2 t_0} \int_{-\infty}^{\infty}dt
\int_{-\infty}^{\infty} dM
e^{-\lambda M /\vp} \phienv(M) \delta(M - \vp t)
\curly {\phicoh(N-M,t) + 2 \phiinc(N-M,t)}
                                                                \end{eq}
This is the contribution to $\phideaddotbar$ due to termination by
combination by one single $\delta$ pulse belonging to the coherent
living MWD, throughout its lifetime.  Combination occurs either with
other coherent chains or by ``cross-coupling'' with chains belonging
to $\phiinc$.  In eq. \eqref{light} we have assumed that chain
transfer is a weak effect, \ie that $\lambda$ is small in a sense to
be quantified below, and we keep the leading correction only.  Thus we
discarded the ${\cal O}(\lambda^2)$ $\phiinc \phiinc$ termination
term.

Now the coherent-coherent coupling term in eq. \eqref{light} is equal to
that of eq. \eqref{statphys}, multiplied by $e^{-\lambda N/\vp}$.  The
only difficulty is the coherent-incoherent cross-term
$\phideadcombcrossdotbar$.  In Appendix E it is shown that
                                          \begin{eqarray}{pelikan-offspring}
\phideadcombcrossdotbar(N) = 
{\lambda e^{-\lambda N /\vp} \over \vp t_0} 
{\kc \over k} (1-\beta)
&& \left\{
\beta^L [(L+1)\beta - L] \Psi_0 + 
\beta^L (L+1)(1-\beta)  \Psi \paren{rN \over 2\vp} + 
\right.
                                        \drop
&& \left.
L \beta^{L-1} (1-\beta) \Psi \paren{rN + n_0 \over 2 \vp} -
\beta^L \Psi \paren{rN \over \vp}
\right\}
\period
                                                        \end{eqarray}

Let us calculate now the disproportionation and chain transfer
contributions, $\phideaddispdotbar$ and $\phideadtransdotbar$.  Time
averaging the last two terms of eq. \eqref{parrot} one has (bar
denotes time average)
                                                \begin{eq}{boat}
\phideaddispdotbar(N) + \phideadtransdotbar(N) = 
\kd \, \overline{\phi(N) \Psi} + \lambda \overline{\phi}(N)
\period
                                                                \end{eq}
Now from eq. \eqref{bag} one has $k \, \overline{\phi(N) \Psi} +
\lambda \overline{\phi}(N) = -\vp {\partial \overline{\phi} / \partial
N}$, using $\overline{\phidot} = 0$ in stationary state.  Thus using eq.
\eqref{boat},
                                                \begin{eq}{bell}
\phideaddispdotbar(N) + \phideadtransdotbar(N) = 
- {\kd \over k} \vp {\partial \overline{\phi} \over \partial N} +
{\kc \over k }\lambda \overline{\phi}(N) \period
                                                                \end{eq}
The rhs of eq. \eqref{bell} is calculated in Appendix F (see eq.
\eqref{hunter}) using the expressions for $\phi$ derived in section
3.1. 

Thus collecting all terms, $\phideaddotbar = \phideadcombdotbar +
\phideaddispdotbar + \phideadtransdotbar$, and noticing that 
the $\kc/ k$ term of eq. \eqref{hunter} cancels the last term
of eq. \eqref{pelikan-offspring}, we have
                                             \begin{eq}{deadtransfer}
\phideaddotbar(N) = e^{-\lambda N /\vp} 
\curly{ \phideadidealdotbar (N) + \varepsilon \, \phideadincdotbar(N)}
\comma
\gap
\varepsilon \equiv {4 \lambda \over (1-\beta) k \Psi_0}
= {4 \lambda \beta t_0 \over (1-\beta)^2}
\comma
                                                              \end{eq}
where
                                                \begin{eqarray}{dead-inc}
\phideadincdotbar(N) \equiv 
C && 
\left[
{\kc \over k}
\left\{
\beta^L [(L+1)\beta - L] + 
\beta^L (L+1)(1-\beta)  f \paren{rN \over 2} 
\right.
\right.
                                        \drop
&& \left.
\left.
+  L \beta^{L-1} (1-\beta) f \paren{rN + n_0 \over 2}
\right\}
\ + \ 2 {\kd \over k} \beta^L f(rN)
\right]
\period
                                                                \end{eqarray}
The two contributions to the dead MWD, $\phidead = t \phideaddotbar$,
are plotted in fig. \ref{dead_trans}.  The incoherent part due
to transfer events has no discontinuities, \ie
$\phideadincdotbar(Ln_0^-) = \phideadincdotbar(L n_0^+)$, as may be
seen using eq. \eqref{f}.  It is proportional to the small
parameter $\varepsilon \approx \lambda t_0$.

Since the only discontinuous term in eq. \eqref{deadtransfer} is
$\phideadidealdotbar$, using eq. \eqref{disc} the discontinuity at $N
= L n_0$ is
                                                \begin{eq}{disc-transfer}
{\Delta \phidead(L n_0) \over \phidead(0)}
=  e^{-\lambda L t_0} \beta^L {2+4\rho \over 1 + 4 \rho/(1-\beta)
+ \varepsilon (1 + 2 \rho)}
\comma
                                                                \end{eq}
where $\Delta \phidead (L n_0) \equiv \phidead(L n_0^+) - \phidead(L
n_0^-)$.


\section{Broadening and the Inflection Point Rule}

The MWD derived in the previous section, eq. \eqref{deadtransfer}, has
ignored (i) Poisson broadening and (ii) broadening due to the limited 
accuracy of GPC measurements.

Consider first Poisson broadening.  The true length distribution of a
chain growing with mean velocity $\vp$ is not $\delta(N - \vp t)$ as
has been assumed in sections 2 and 3, but rather the Poisson
distribution (see fig. \ref{broad}) which for large $N$ tends to a Gaussian:
                                                \begin{eq}{poisson}
P_t(N) = {(\vp t)^N \over N!} e^{- \vp t}
\approx c e^{-(N - \vp t)^2/(2 \vp t)} 
\comma \gap
(N \gg 1) 
\period      
                                                                \end{eq}
Here $c = (2 \pi \vp t)^{-1/2}$ is a normalization constant.

Now since the termination rate constant $k$ is independent of chain
length, clearly broadening leaves the total number of living chains
$\Psi(t)$ unchanged.  Hence the field $H(t)$ is unchanged, and thus a
group of living chains injected at $t=0$ are still depleted in number
by the factor $e^{- \int_0^t H}$ after time $t$, regardless of which
chains are a little longer than $\vp t$ and which a little shorter.

Moreover we show in Appendix G that even in the presence of chain
transfer, the Poisson broadened living MWD, $\phibroad(N)$, can be
expressed in terms of the unbroadened living MWD, $\phi(N)$,
calculated in the previous sections as follows:
                                                \begin{eq}{pirate}
\phibroad(N,t) = \int_{-\infty}^{\infty} dM \phi(M,t) \Delta(N;M)
\comma
\gap
\Delta(N; \Nbar) \equiv \inverse{\sqrt{2 \pi \Nbar}} e^{- (N - \Nbar)^2/(2 \Nbar)}
                                                                \end{eq}
Thus the $\delta$-pulses of $\phi$ in eq. \eqref{living-pure} (or of
$\phicoh$ in eq. \eqref{phicoh} if chain transfer is present) are
replaced by $\Delta$'s, namely Gaussians of mean $\Nbar$ and width
$\sqrt{\Nbar}$.  Note that with increasing $\Nbar$, the relative width
of the broadened $\Delta$-functions decreases as $1/\sqrt{\Nbar}$.

Then substituting eq. \eqref{pirate} into eq. \eqref{parrot} we find
that the dead MWD, $\phideadbroad$, is also obtained by integrating
the dead MWD in the absence of Poisson broadening, eq. \eqref{deadtransfer},
against a $\Delta$ function:
                                                \begin{eq}{convo}
\phideadbroad (N)
= \int_{-\infty}^{\infty} dM\, \phidead(M) \Delta(N;M) 
\period
                                                                \end{eq}
(We used the identity $\int_{-\infty}^{\infty} dM \Delta(M;P)
\Delta(N-M;L) = \Delta(N;L+P)$ for the termination by combination term
of eq. \eqref{parrot}).  Here $\phidead = t \phideaddotbar$.  An
example of the broadened MWD is shown in fig. \ref{dead_broad}, where
$\phideadbroad$ is calculated by numerical integration of eq.
\eqref{convo}.  Eq. \eqref{convo} has been called ``a posteriori
Poisson broadening'' \cite{kornherr:olaj_plp_revisited}. In the
present work we proved its validity.

Now the accuracy with which the resulting dead MWD of eq.
\eqref{convo} is measured is limited by the resolution of the GPC
equipment.  Thus the MWD implied by the GPC output reading is
                                                \begin{eq}{distort}
\phideadread(N) = \int_{-\infty}^{\infty} dM\, \phideadbroad(M) \, G(N;M)
\comma
                                                                \end{eq}
where $G(N;M)$ is the experimental reading (normalized to unity) given a
perfectly monodisperse MWD input of chains of length $M$.

Consider values of $N$ near the discontinuity of the unbroadened dead
MWD, $N = L n_0$, and let us examine the location of the inflection
point.  In the following we will assume that near the discontinuity to
leading order $G$ can be approximated by a Gaussian distribution,
$G(N;n) = (2 \pi \sigma^2)^{-1/2} e^{-(N-n)^2/(2
\sigma^2)}$, where $\sigma$ is a measure of the experimental
resolution near $L n_0$.  Substituting $\phideadbroad$ from eq.
\eqref{convo} in eq. \eqref{distort} one has
                                                \begin{eq}{cage}
\phideadread(N) = \int_{-\infty}^{\infty} dM\, \phidead(M) 
\Deltatilde(N;M)
\comma \gap
\Deltatilde(N,M) \equiv {e^{-(N-M)^2/[2(M+\sigma^2)]} \over 
{[2 \pi (M + \sigma^2)]}^{1/2}}
\comma
                                                                \end{eq}
after using $\int_{-\infty}^{\infty} dP\, \Delta (P;M) G(N;P) =
\Deltatilde (N;M)$ for Gaussian $G$.

In Appendix H the curvature of $\phideadread$ is evaluated close to a
discontinuity (eq. \eqref{olympia}). Setting this to zero one finds
two roots, one irrelevant at $\delta N \equiv N - L n_0 \approx 2(L
n_0 + \sigma^2)$, and one corresponding to the point of inflection at
                                                \begin{eq}{root}
\delta N \approx 
\curly{
{\phi_+' - \phi_-' \over \phi_+ - \phi_-} (L n_0 + \sigma^2)
 - {1 \over 2}
} \comma  
\ \ \ \
\phi_{+/-} \equiv \phideadideal(Ln_0^{+/-}) \comma \ \ \ \
\phi_{+/-}' \equiv {d\phideadideal \over dN}(Ln_0^{+/-}) \comma 
                                                                \end{eq}
Now in PLP experiments, parameters are normally chosen such that the
dead MWD decreases by a factor of order unity between successive
peaks.  Hence $\phi_+'$ is of order $\phi_+/n_0$ and similarly for
$\phi_-$.  Since $\phi_+' < \phi_-' < 0$ and $\phi_+ > \phi_- > 0$,
and taking $L = {\cal O}(1)$, one has from eq. \eqref{root},
                                                \begin{eq}{displace}
\delta N \approx -A \, (1 + B\, \sigma^2/ n_0) \comma
                                                                \end{eq}
where $A,B > 0$ are constants of order unity.  This theoretical result
is valid provided $n_0^{-1/2}\ll 1$ and $\beta$ is of order unity.
Note that $\delta N$ is always negative.

Consider first very good experimental resolution, $\sigma \ll
n_0^{1/2}$.  In this case the observed broadening is predominantly
Poissonian.  According to eq. \eqref{displace}, the displacement of
the inflection point of the measured MWD away from $Ln_0$ is then of
order one monomer.  This is an unimportant correction which is of the
same order as the accuracy of the continuum description of the MWD
employed in our analysis.  We conclude that for this case the
inflection point is the appropriate point from which to extract $\kp$.

Now when $\sigma \gg n_0^{1/2}$, the broadening of the measured MWD is
dominated by the finite experimental resolution.  In this case the
inflection point is of order $\sigma^2/n_0 \gg 1$ monomers away from
$L n_0$.  Therefore for large $\sigma$, using the location of the
inflection point to deduce $n_0$ leads to a systematic underestimation
of $\vp$.  We remark however that this is more accurate than using the
local minimum or maximum of the MWD which will then be at a distance
of order $\sigma$ away from $n_0$ (since we have assumed that $\sigma
\ll n_0$, hence $\sigma \gg \sigma^2/n_0$).

As a specific example, consider the experimental MWD of fig.
\ref{hutchmwd}.  In this figure, the distance between the local
minimum and maximum of the MWD around the inflection point near
molecular weight 50,000 ($n_0 \approx 500$) is of order 200 monomers.
In this case Poisson broadening would only account for a width
$\approx n_0^{1/2} \approx 22$ monomers.  Hence broadening appears to
be mainly due to limited experimental accuracy and $\sigma \approx
100$.  The relative error using the inflection point for the
calculation of $\vp$ would then lead to an error of order
$\sigma^2/n_0^2 \approx 4$\%.


\section{Conclusions}

In this work, we have calculated analytically the number MWD of dead
chains produced by pulsed laser polymerization at low conversions.
The following effects have been included: living chain termination by
both combination and disproportionation; chain transfer of living
chain radicals to the environment; broadening effects.  Ignoring
broadening, we have found the MWD is given by
                                                \begin{eq}{conclusion}
\phidead(N) =  e^{-\lambda N /\vp} 
\curly{ \phideadideal (N) + \varepsilon \, \phideadinc(N)}
\comma
\gap
\varepsilon = {4 \lambda \beta t_0 \over (1-\beta)^2}
\comma
                                                                \end{eq}
where
                                                \begin{eqarray}{conclusion-2}
\phideadideal(N) = 
\const 
&&
\square{
{\kc \over k }
\curly{
(L+1) \beta^L f^2\paren{rN \over 2} + L \beta^{L-1} 
f^2\paren{rN + n_0 \over 2} 
}
+ 
4 {\kd \over k} 
\curly{{\beta^L \over 1-\beta} f^2(rN)}
}
\comma
                                        \ddrop
\phideadinc(N) \equiv
\const && 
\left[
{\kc \over k}
\left\{
\beta^L [(L+1)\beta - L] + 
\beta^L (L+1)(1-\beta)  f \paren{rN \over 2} 
\right.
\right.
                                        \drop
&& \left.
\left.
+  L \beta^{L-1} (1-\beta) f \paren{rN + n_0 \over 2}
\right\}
\ + \ 2{\kd \over k} \beta^L f(rN)
\right]
\comma
                                        \drop
f(n) \equiv && 
{\beta  \over \beta+(1-\beta) n/n_0} 
\period
                                                                \end{eqarray}
Here $N \equiv L n_0 + rN$, and ``$\const$'' is a normalization constant.
The dead MWD is discontinuous at integer multiples of $n_0$ and is a
function of the following independent parameters: $n_0$, $\lambda
t_0$, $\kd/k$, $\kc/k$, and $\beta$.  Here $\beta$ (see eq.
\eqref{fact}), the fraction of living chains surviving between 2
successive pulses, is a function of the dimensionless parameter $k
\Ribar t_0^2$, where $\Ribar t_0$ is the concentration of new radicals
produced per pulse.  Eq. \eqref{conclusion} applies in the limit in
which chain transfer is weak, \ie $\lambda t_0 \ll 1$. Setting
$\kc=0$, eq. \eqref{conclusion} reduces to the dead MWD derived in
refs.
\cite{ziffererolaj:plp:chaintrans1,ziffererolaj:plp:chaintrans2},
while setting $\lambda=0$ one recovers the dead MWD derived in refs.
\cite{olaj:plp,kornherr:olaj_plp_revisited}.

An important quantity in eqs. \eqref{conclusion} and
\eqref{conclusion-2} is the relative magnitude of the dead MWD
discontinuities, shown in eq. \eqref{disc-transfer}.  Optimally large
discontinuities typically correspond to values of $\beta$ intermediate
between zero and unity as has been thoroughly examined by numerical
simulations \cite{buback:plp_simul,hutchinson:plp:vinyl_acetate}
and discussed in ref. \cite{beuermann:plp_beta_limits}.  For $\beta
\ll 1$ these relative magnitudes are undesirably small.  The limit
$\beta \gt 1$ (realised when $k \Ribar t_0^2$ is small) may be
undesirable also, but this depends on the mode of termination: (i) For
the extreme of termination by disproportionation only ($\rho \gt
\infty$ in eq. \eqref{disc-transfer}) the discontinuities vanish in
this limit while (ii) for termination by combination only ($\rho = 0$)
the discontinuities attain their maximum value as $\beta \gt 1$.  Thus
if combination is strongly dominant, by choosing small $\Ri$ this may
be a desirable limit (though noise effects are then particularly
strong since the absolute amplitude of the dead MWD is very small).
In a typical case though where termination occurs by both combination
and disproportionation the optimal choice corresponds to intermediate
$\beta$ values.

The effect of chain transfer is to reduce the amplitude of the entire
MWD.  The amplitude of the discontinuities is reduced by a factor
$e^{-\lambda t_0 N/n_0}$.  (It follows that if $\lambda t_0 > 1$, most
living chains suffer a chain transfer event before growing to a length
of order $n_0$; consequently the MWD amplitude at $N=n_0$ is small.
Under such conditions PLP is not a useful method for measuring $n_0$.)
In addition to these effects, transfer introduces an extra
``incoherent'' contribution to the dead MWD (see fig.
\ref{dead_trans}) which has no discontinuities and whose magnitude is
a fraction $\lambda t_0$ of the dead MWD.

We have also treated 2 types of broadening: Poisson and distortion by
MWD measuring equipment.  Poisson broadening smoothes out the
discontinuities of the dead MWD of eq. \eqref{conclusion}.  The
broadened dead MWD can be computed by performing numerically the
integral of eq. \eqref{convo}.  We found that for $\beta$ of order
unity the position of the inflection point of the smooth MWD is not
shifted significantly away from the initial discontinuity.  ( For very
small $\beta$ the inflection point rule leads to underestimation of
$n_0$.)  Thus, if Poisson broadening dominates, then its location can
be used reliably to infer $\kp$.

In practice, the 2nd type of broadening, due to limited GPC
resolution, may swamp the Poisson effect.  We found that when the band
width $\sigma$ of the GPC output reading, given a perfectly
monodisperse sample of degree of polymerization $L n_0$, is much
greater than $n_0^{1/2}$, experimental broadening then dominates near
the discontinuity at $L n_0$.  This leads to a displacement of the
point of inflection by approximately $\sigma^2/n_0$ monomers below $L
n_0$.  We remark however that (for $\beta$ of order unity and $\sigma
\ll n_0$) the point of inflection is closer to $n_0$ than the position
of either the minimum or maximum peaks.

\vi

{\bf Acknowledgments.} This work was supported by the National Science
Foundation under grant no. DMR-9816374.  We thank Erdem Karatekin for
illuminating discussions.


\def\appendix{\par\clearpage
  \setcounter{section}{0}
  \setcounter{subsection}{0}
  \@addtoreset{equation}{section}
  \def\theequation{\thesection\arabic{equation}}
  \def\thesection{\Alph{section}}
  \def\thesubsection{\arabic{subsection}}}

{\appendix

\section{Solution to Living Chain Dynamics for Arbitrary $\Ri(t)$}

In this appendix we solve eq. \eqref{phi} for general radical
production rate $\Ri(t)$.  Defining $\xi \equiv t-N/\vp$ and $f(\xi,t)
\equiv \phi(N,t)$ we have from eq. \eqref{phi}
                                                \begin{eq}{bank}
\left. {\partial f \over \partial t} \right|_\xi = 
- H(t) f + \delta (\vp t - \vp \xi) \Ri(t) \comma
                                                                \end{eq}
with $\phi(N<0,t) = 0$ implying $f(\xi>t) = 0$.  The solution of eq.
\eqref{bank} is 
                                                \begin{eq}{bird}
f(\xi,t) 
= 
\inverse{\vp} \int_{-\infty}^{t} dt' e^{-\int_{t'}^{t} H}
\delta(t'-\xi) \Ri(t')
= 
\Theta(t-\xi) \inverse{\vp} e^{-\int_\xi^t H} \Ri(\xi) \comma
                                                                \end{eq}
where $\Theta$ is the step function, $\Theta(x>0)=1$ and  $\Theta(x<0) =
0$.  Hence
                                                \begin{eq}{phisoln}
\phi(N,t) = {\Theta(N) \over \vp} e^{- \int_{t-N/\vp}^{t} H(t') dt'} \Ri(t-N/\vp)
\period
                                                                \end{eq}


\section{Properties of the Reaction Field $H(t)$}

Since $\Psidot = - H \Psi$ within one cycle (see eq.
\eqref{psidynamics}), it follows that if both $t_1$ and $t_2$ belong
to the same cycle then
                                                \begin{eq}{zap}
e^{-\int_{t_1}^{t_2} H(t') dt'} = {\Psi(t_2) / \Psi(t_1)}
\comma \gap (\ [t_1] = [t_2]\ ) \period
                                                                \end{eq}
Since $\Psi(t)$ drops by a factor $\beta$ in one cycle of duration
$t_0$ we also have
                                                \begin{eq}{hproperty}
e^{-\int_{t_1}^{t_2} H(t') dt'} = 
e^{-\int_{[t_1] t_0}^{[t_2] t_0} H(t')dt'}
e^{-\int_{rt_1}^{rt_2} H(t')dt'} = 
{\Psi(t_2) \over \Psi(t_1)} \beta^{[t_2]-[t_1]}
                                                                \end{eq}
after using eq. \eqref{zap}.

\section{General Expressions for Dead MWD's Generated by Combination}

Consider two living MWDs $\phi_1, \phi_2$ which are periodic in time
with period $t_0$ such that $\phi_1$ can be written as
                                                \begin{eq}{phione}
\phi_1(N,t) = \sum_{j=-\infty}^{\infty} A_1 (N) f_1(N - \vp t - j
n_0)
\comma
                                                                \end{eq}
where $f_1$ has period $n_0$.  Suppose dead chains are generated only
by bimolecular reactions involving one 1 and one 2 chain.  If only
combination reactions occur, the time averaged rate of dead chain
generation then satisfies
                                                \begin{eqarray}{equivalence}
\phideaddotbar(N) &=& {k \over t_0} 
\int_0^{t_0} dt
\int_{-\infty}^{\infty} dM \, \phi_1(M,t) \phi_2(N-M,t)
                                                \drop
&=&
{k \over t_0}
\sum_{j=-\infty}^{\infty}
\int_0^{t_0} dt
\int_{-\infty}^{\infty} dM
A_1 (M) f_1(M - \vp t - j n_0)
\phi_2(N-M,t) 
                                       \drop
&=& {k \over t_0}
\sum_{j=-\infty}^{\infty}
\int_{ j t_0}^{(j+1)t_0} dt'
\int_{-\infty}^{\infty} dM
A_1 (M) f_1(M - \vp t')
\phi_2(N-M,t' + j t_0) 
\comma \drop
                                                                \end{eqarray}
where we changed variables to $\vp t' = \vp t + j n_0$.  We remark
that when there is only one chain type reacting with its own type to
produce dead chains, then $\phi_1 = \phi_2$ and the rhs of eq.
\eqref{equivalence} should be divided by 2.
Since $\phi_2$ has period $t_0$ one has the following equivalent
expression:
                                                \begin{eqarray}{mean}
\phideaddotbar(N) = {k \over t_0}
\int_{-\infty}^{\infty} dt'
\int_{-\infty}^{\infty} dM \,
A_1 (M) f_1(M - \vp t')
\phi_2(N-M,t') 
\period
                                                                \end{eqarray}
With $A_1(N,t) \gt \phienv(N)$, $f_1 \gt \delta$, $\phi_2 \gt \phi_1$,
$k \gt \kc$ and dividing by 2 this yields eq. \eqref{statphys} of the
main text.


\section{Ideal Case: Combination Part of Dead Chain MWD, eq. (17)}

Changing variables $n = 2 \vp t + j n_0$ in eq. \eqref{statphys} and using eq.
\eqref{living-pure} to substitute for $\phi$ in eq. \eqref{statphys} one has 
                                                \begin{eqarray}{routine}
\phideadcombdotbar(N) 
&=& {\kc \over 4 \vp t_0}
\int_{-\infty}^{\infty} dn \sum_{j=-\infty}^{\infty} 
\phienv \paren{n + j n_0 \over 2} \phienv \paren{n - j n_0 \over 2} \delta (N - n)      
                                        \drop
&=& {\kc \over 4 \vp t_0} \sum _{j=-\infty}^{\infty} 
\phienv \paren{N + j n_0 \over 2} \phienv \paren{N - j n_0 \over 2}
\period              
                                                          \end{eqarray}

Writing $N = L n_0 + rN$, with $L \equiv [N]$, we evaluate the sum of
eq. \eqref{routine} by noting that since $\phienv$ is zero for negative values
of its arguments, only $-L \le j \le L$ contribute.  Now using eq.
\eqref{a} one has
                                                \begin{eqarray}{mouse}
\phienv \paren{(L+j) n_0 + rN \over 2} \phienv \paren{(L-j) n_0 + rN \over 2}
= 
\casesbracketsii
{ \beta^L \phienv^2(rN/2)}              {L-j \ \mbox{even}}
{ \beta^{L-1} \phienv^2((rN+n_0)/2)}    {L-j \ \mbox{odd}}
                                                        \ddrop
(-L \le j \le L)
\drop
                                                                \end{eqarray}
Noting that the number of even and odd $L-j$ values in the interval
$-L \le j \le L$ is $L+1$ and $L$, respectively, and using eqs.
\eqref{mouse} and \eqref{routine}, eq. \eqref{rat} of the main text is
derived after expressing $\phienv$ in terms of $\Psi$ using eq. \eqref{a}.


\section{Derivation of Coherent-Incoherent Coupling Term
$\overline{\dot{\phi}}_{\rm dead,cross}^{\rm comb}$}

Using eqs. \eqref{phicoh} and eq. \eqref{a}, one has from eq.
\eqref{light} after integrating over the $\delta$-function:
                                                \begin{eqarray}{criss}
\phideadcombcrossdotbar(N) &=&
{\kc \over \vp t_0}
\int_0^N dM \, \phienv(M) 
\curly{
\lambda  e^{-\lambda N /\vp} \Psi(M/\vp) \beta^{1+[N-2M] + [M]}
}
                                \drop
&=& {\lambda \kc e^{-\lambda N /\vp} \over \vp t_0} (1-\beta)
\int_0^N dM \, \Psi^2(M/\vp) \beta^{L+1 + [rN - 2rM]} 
                                \ddrop
N &\equiv& L n_0 + rN \comma \gap M \equiv P n_0 + rM
\period
                                                                \end{eqarray}
Here we used the identity $-[N-2M] = [2M - N] +1$.
The integrand of eq. \eqref{criss} is independent of $P$ and only
depends on the relative magnitude of $rN$ with respect to $rM$.  Since
$O \le P \le L$ we may split $\int_0^N$ into $L$ integrals
$\int_0^{rN/2} + \int_{rN/2}^{(rN+n_0)/2} + \int_{(rN + n_0)/2}^{n_0}$
corresponding to $P \le L-1$, plus the $P=L$ term whose upper
integration limit is $rN$ instead of $n_0$:
                                                \begin{eqarray}{pelikan}
\phideadcombcrossdotbar(N) = 
&& {\lambda \kc e^{-\lambda N /\vp} \over \vp t_0} (1-\beta)\times
                                \drop
&& \left\{ 
     (L+1) \beta^{L+1} \int_0^{rN/2} d(rM) \Psi^2(rM/\vp) 
+    (L+1) \beta^{L} \int_{rN/2}^{(rN+n_0)/2} d(rM) \Psi^2(rM/\vp)
\right. 
                                                \drop
&& \left. 
+    L \beta^{L-1} \int_{(rN+n_0)/2}^{n_0} d(rM) \Psi^2(rM/\vp) 
+    \beta^L \int_{(rN+n_0)/2}^{rN} d(rM) \Psi^2(rM/\vp) 
\right\}
                                                            \end{eqarray}
The final expression for $\phideadcombcrossdotbar$, eq.
\eqref{pelikan-offspring} of the main text, follows from eq.
\eqref{pelikan} after using the identity $- k \int_{t_1}^{t_2} dt
\Psi^2 =
\Psi(t_2) - \Psi(t_1)$ for $0 < t_1,t_2 < t_0$.

\section{Disproportionation and Chain Transfer Contribution to Dead MWD}

From eq. \eqref{phicoh}, the time average of $\phicoh$ is 
                                                \begin{eq}{coh-mean}
\overline{\phi}_{\small coh}(N) = {1 \over \vp t_0} e^{- \lambda N /\vp} 
\phienv(N) \period
                                                                \end{eq}
Hence using the definition of $\phienv$  (eq. \eqref{a}) and $\Psidot(t) = - k
\Psi^2(t)$ for $0 < t < t_0$,
                                                \begin{eq}{basket}
- \vp {\partial \overline{\phi}_{\small coh} \over \partial N}
=
{1 \over \vp t_0} e^{- \lambda N/\vp} \phienv(N)\,
\square{\lambda + k \Psi(rN/\vp)}
                                                                \end{eq}

The time average of $\phiinc$ is calculated after integrating the 3rd
expression in eq. \eqref{phicoh} from 0 to $t_0$.  The value of the
integrand depends on the relative magnitudes of $rN/\vp$ and $t$:
                                                \begin{eq}{dip}
\overline{\phi}_{\small inc}(N) =
{\lambda e^{- \lambda N /\vp} \over \vp t_0}
\beta^{[N]+1}
\curly{
\int_{0}^{rN/\vp} dt\,\Psi(t) +
\inverse{\beta} \int_{rN/\vp}^{t_0} dt\, \Psi(t) 
}
\period
                                                                \end{eq}
Hence neglecting terms proportional to $\lambda^2$,
                                                \begin{eq}{yellow}
- \vp {\partial \overline{\phi}_{\small inc} \over \partial N}
=
\inverse{\vp t_0} \lambda e^{- \lambda N /\vp} \phienv(N)
+ 
{\cal O}(\lambda^2)
\period
                                                                \end{eq}
Thus from eqs. \eqref{coh-mean}-\eqref{yellow} one has 
                                                \begin{eq}{hunter}
- {\kd \over k} \vp {\partial \overline{\phi} \over \partial N} +
{\kc \over k }\lambda \overline{\phi}(N)
= 
C e^{-\lambda N /\vp}
\curly{
4{\kd \over k}{\beta^{[N]} \over 1-\beta} f^2(rN)
+
\varepsilon \paren{2 {\kd \over k} + {\kc \over k}} \beta^{[N]} f(rN)
} \comma
                                                                \end{eq}
after using eqs. \eqref{a} and \eqref{f} and the definition of $\varepsilon$
in eq. \eqref{deadtransfer}.  Here we neglected ${\cal O}(\lambda^2)$
terms. 


\section{Poisson Broadening of Living MWD: Derivation of eq. (33)}

In order to include broadening effects in the living dynamics, a
``diffusion'' term must be added in eqs. \eqref{phi} and \eqref{bag}
as follows:
                                                \begin{eq}{pde}
\phibroaddot = -\vp {\partial \phibroad \over \partial N}
+ {\vp \over 2} {\partial^2 \phibroad \over \partial N^2}
 - \Htilde (t) \phi +  \Ritilde(t)  \delta(N) \comma
                                                                \end{eq}
where $\Htilde(t) \equiv H(t) + \lambda$ and $\Ritilde \equiv \Ri(t) +
\lambda \Psi(t)$.  In the absence of the last
two sink and source terms, eq. \eqref{pde} describes the evolution of
the probability distribution function a unidirectional walk with step
size unity and velocity $\vp$, in the continuum limit.  The diffusion
coefficient for such a walk \cite{reif:book} is $\vp/2$.

Now in the absence of the source and sink terms, the propagator of eq.
\eqref{pde} for a pulse generated at $N=0$ at $t=0$,
is simply $\Delta(N;\vp t)$.  (Here we neglected the condition
$\phibroad(N<0;t) = 0$ which implies that the propagator be zero for
negative $N$ values.  However for $\vp t > 1$, the amplitude of
$\Delta(N;\vp t)$ becomes exponentially small for negative $N$.  This
implies that the use of $\Delta$ is a very good approximation provided
we examine the living MWD for $N \gg 1$.)  Hence the solution of eq.
\eqref{pde} is
                                                \begin{eq}{broad-solution}
\phibroad(N,t) = 
 \int_{-\infty}^{t} dt' e^{-\int_{t'}^{t} \Htilde} \Ri(t') \,
\Delta\paren{N;\vp(t-t')} \period
                                                                \end{eq}
This solution is identical to the solution of the
unbroadened living MWD in eq. \eqref{bird}, after replacing
$\delta(t'-\xi) = \vp \delta(N-\vp(t-t'))$ in eq. \eqref{bird} by
$\Delta(N;\vp(t-t'))$.  Hence eq. \eqref{pirate} follows immediately
as one may check after replacing $\phi$ from eq. \eqref{bird} into
eq. \eqref{pirate} and checking that eq. \eqref{broad-solution} is
recovered.


\section{Curvature of Broadened Dead MWD Near the Point of Inflection}

Considering molecular weights near $L n_0$, expanding $\phidead(n)$
one has (see eq. \eqref{root} for notation)
                                                \begin{eq}{pure-expand}
\phidead (n) = \paren{ \phi_+ + \phi_+' \delta n + ...}
\Theta(\delta n)
+ \paren{ \phi_- + \phi_-' \delta n + ...}
\Theta(-\delta n)
\comma \gap
\delta n \equiv n - L n_0 \ \period 
                                                                \end{eq}
 Since the amplitude of the unbroadened dead MWD decreases smoothly by
a factor of order unity between successive peaks, the magnitude of
$\phi_+'$ is of order $\phi_+/ n_0$, and similarly for $\phi_-'$.  Now
since the $\delta n$ which will survive in the integration of eq.
\eqref{convo} around $N \approx L n_0$ is of order $\Sigma \equiv (L
n_0 + \sigma^2)^{1/2}$, eq. \eqref{pure-expand} is an expansion up to
order $\Sigma/n_0$.  Approximating $\Deltatilde$ with a Gaussian of
fixed width, and keeping terms up to order $1/\Sigma$ (note $\Sigma
/n_0 \ge 1/\Sigma$):
                                                \begin{eqarray}{del-expand}
\Deltatilde(N; n)
&=& 
\inverse{(2 \pi \Sigma^2)^{1/2}} e^{-(\delta N - \delta n)^2/(2 \Sigma^2)}
\curly{
1 - {\delta n \over 2 \Sigma^2}
+
{\delta n (\delta N - \delta n)^2 \over 2 \Sigma^4 } + 
...
} \comma
                                        \ddrop
\delta N &\equiv& N - L n_0 \period
                                                                \end{eqarray}
Substituting eqs. \eqref{pure-expand} and \eqref{del-expand} in eq.
\eqref{convo} we obtain the following approximation to the
MWD output reading near $L n_0$:
                                                \begin{eqarray}{tea}
\phideadread(N) 
&=& 
\inverse{(2 \pi \Sigma^2)^{1/2}}
\int_0^{\infty} d(\delta n)
\curly{
\phi_+ + 
\delta n 
\paren{
\phi_+' - {\phi_+ \over 2 \Sigma^2} + 
{\phi_+(\delta N - \delta n)^2 \over 2 \Sigma^4}
}
}
e^{-(\delta N - \delta n)^2/(2 \Sigma^2)}  
                                                \ddrop
&+&
\inverse{(2 \pi \Sigma^2)^{1/2}} \int_{-\infty}^0 d(\delta n)
\curly{
...
}_{\phi_+,\phi_+' \gt \phi_-, \phi_-'}        
                                                                \end{eqarray}
A straightforward differentiation of eq. \eqref{tea} gives 
                                                \begin{eq}{olympia}
{d^2 \phideadread \over dN^2}
=
\inverse{(2 \pi \Sigma^2)^{1/2}}
\curly{
\phi_+' - \phi_-' +
{\phi_- - \phi_+ \over 2 \Sigma^2}
\square{
1 + 2 \delta N - {(\delta N)^2 \over \Sigma^2} 
}
}
e^{-(\delta N )^2/(2 \Sigma^2)}  \period
                                                                \end{eq}


}

\pagebreak


\begin{thebibliography}{10}

\bibitem{flory:book}
P.~Flory, {\it Principles of Polymer Chemistry}, Cornell University Press,
  Ithaca, New York 1971.

\bibitem{odian:book}
G.~Odian, {\it Principles of Polymerization}, John Wiley and Sons, New York
  1981.

\bibitem{iupac:frp_3}
M.~Buback, , R.~G. Gilbert, R.~A. Hutchinson, B.~Klumperman, F.-D. Kuchta,
  B.~Manders, K.~F. O'Driscoll, G.~T. Russell, J.~Schweer, {\it Macromol. Chem.
  Phys.} {\bf 1995}, {\it 196}, 3267--3280.

\bibitem{beuermannbuback:plp_review_2002}
S.~Beueremann, M.~Buback, {\it Progress in Polymer Science} {\bf 2002}, {\it
  27}, 191--254.

\bibitem{coote:plp_review}
M.~L. Coote, M.~D. Zammit, T.~P. Davis, {\it Trends in Polymer Science} {\bf
  1996}, {\it 4}, 189--196.

\bibitem{olaj:plp}
O.~F. Olaj, I.~Bitai, F.~Hinkelmann, {\it Makromol. Chem.} {\bf 1987}, {\it
  188}, 1689--1702.

\bibitem{hutchinson:plp:methacrylates}
R.~A. Hutchinson, J.~D.~A.~Paquet, J.~H. McMinn, R.~E. Fuller, {\it
  Macromolecules} {\bf 1995}, {\it 28}, 4023--4028.

\bibitem{hutchinson:plp:methacrylates2}
R.~A. Hutchinson, S.~Beuermann, D.~A. {Paquet Jr.}, J.~H. McMinn, C.~Jackson,
  {\it Macromolecules} {\bf 1998}, {\it 31}, 1542--1547.

\bibitem{hutchinson:plp:alkylmethacrylates}
R.~A. Hutchinson, S.~Beuermann, D.~A. {Paquet Jr.}, J.~H. McMinn, {\it
  Macromolecules} {\bf 1997}, {\it 30}, 3490--3493.

\bibitem{hutchinson:plp}
R.~A. Hutchinson, M.~T. Aronson, J.~R. Richards, {\it Macromolecules} {\bf
  1993}, {\it 26}, 6410--6415.

\bibitem{hutchinson:plp:vinyl_acetate}
R.~A. Hutchinson, J.~R. Richards, M.~T. Aronson, {\it Macromolecules} {\bf
  1994}, {\it 27}, 4530--4537.

\bibitem{shipp:plp:methacrylonitrile}
D.~A. Shipp, T.~A. Smith, D.~H. Solomon, G.~Moad, {\it Macromol. Rapid Commun.}
  {\bf 1995}, {\it 16}, 837--844.

\bibitem{hutchinson:plp:chaintransfer}
R.~A. Hutchinson, J.~D.~A.~Paquet, J.~H. McMinn, {\it Macromolecules} {\bf
  1995}, {\it 28}, 5655--5663.

\bibitem{ziffererolaj:plp:thermalinit}
G.~Zifferer, O.~F. Olaj, {\it Macromol. Theo. Simul.} {\bf 1998}, {\it 7},
  157--169.

\bibitem{sarneckischweer:plp}
J.~Sarnecki, J.~Schweer, {\it Macromolecules} {\bf 1995}, {\it 28}, 4080--4088.

\bibitem{buschwahl:plp_transfer}
M.~Busch, A.~Wahl, {\it Macromol. Theory Simul.} {\bf 1998}, {\it 7}, 217--224.

\bibitem{buback:plp_simul}
M.~Buback, M.~Busch, R.~A. Lammel, {\it Macromol. Theory Simul.} {\bf 1996},
  {\it 5}, 845--861.

\bibitem{odriscollkuindersma:plp_montecarlo}
K.~F. O'Driscoll, M.~E. Kuindersma, {\it Macromol. Theory Simul.} {\bf 1994},
  {\it 3}, 469--478.

\bibitem{yan:plp_ch_transfer_numerical}
D.~Yan, M.~Zhang, J.~Schweer, {\it Macromolecules} {\bf 1996}, {\it 29},
  3793--3799.

\bibitem{alexandrov:plp}
A.~P. Alexandrov, V.~N. G. M.~S. Kitai, I.~M. Smirnova, V.~V. Sokolov, {\it
  Sov. J. Quantum Electron. (Engl. Transl.)} {\bf 1977}, {\it 7}(5), 547--550.

\bibitem{kornherr:olaj_plp_revisited}
A.~Kornherr, G.~Zifferer, O.~F. Olaj, {\it Macromol. Theory Simul.} {\bf 1999},
  {\it 8}, 260--271.

\bibitem{evseevnikitin:plp_arbitrary_sequence}
A.~V. Evseev, A.~N. Nikitin, {\it Laser Chem.} {\bf 1995}, {\it 16}, 83--99.

\bibitem{ziffererolaj:plp:chaintrans1}
G.~Zifferer, O.~F. Olaj, {\it Macromol. Theory Simul.} {\bf 1996}, {\it 5},
  901--921.

\bibitem{ziffererolaj:plp:chaintrans2}
G.~Zifferer, O.~F. Olaj, {\it Macromol. Theory Simul.} {\bf 1996}, {\it 5},
  923--938.

\bibitem{ben:interdil_letter}
B.~Friedman, B.~O'Shaughnessy, {\it Europhys. Lett.} {\bf 1993}, {\it 23},
  667--672.

\bibitem{ben:interdil}
B.~Friedman, B.~O'Shaughnessy, {\it Macromolecules} {\bf 1993}, {\it 26},
  5726--5739.

\bibitem{ben:review_ijmp}
B.~Friedman, B.~O'Shaughnessy, {\it Int. J. Mod. Phys. B} {\bf 1994}, {\it 8},
  2555--2591.

\bibitem{odriscollmahabadi:sip}
K.~F. O'Driscoll, H.~K. Mahabadi, {\it J. Polym. Sci.: Polym. Chem. Ed.} {\bf
  1976}, {\it 14}, 869--881.

\bibitem{olaj:plp:kt2}
O.~F. Olaj, A.~Kornherr, G.~Zifferer, {\it Macromol. Rapid Commun.} {\bf 1997},
  {\it 18}, 997--1007.

\bibitem{olaj:plp:kt3}
O.~F. Olaj, A.~Kornherr, G.~Zifferer, {\it Macromol. Rapid Commun.} {\bf 1998},
  {\it 19}, 89--96.

\bibitem{olaj:poisson_plus_kt}
O.~F. Olaj, G.~Zifferer, A.~Korhnherr, {\it Macromol. Theory Simul.} {\bf
  1998}, {\it 7}, 321--326.

\bibitem{olaj:kt4}
O.~F. Olaj, A.~Korhnherr, G.~Zifferer, {\it Macromol. Theory Simul.} {\bf
  1998}, {\it 7}, 501--508.

\bibitem{nikitinevseev:kt}
A.~N. Nikitin, A.~V. Evseev, {\it Macromol. Theory Simul.} {\bf 1997}, {\it 6},
  1191--1210.

\bibitem{nikitinevseev:kt_kp}
A.~N. Nikitin, A.~V. Evseev, {\it Macromol. Theory Simul.} {\bf 1999}, {\it 8},
  296--308.

\bibitem{beuermann:plp_beta_limits}
S.~Beuermann, D.~A. Paquet~Jr., J.~H. McMinn, R.~A. Hutchinson, {\it
  Macromolecules} {\bf 1996}, {\it 29}, 4206--4215.

\bibitem{reif:book}
F.~Reif, {\it Fundamentals of Statistical and Thermal Physics}, McGraw-Hill,
  New York 1965.

\end{thebibliography}


                     \begin{thefigures}{99}

\figitem{propagation}

Schematic of monomer addition to the free radical end of a living
chain.

\figitem{termination}

Termination reaction of 2 living chains of length $N$ and $M$.  Combination
results into one dead chain of length $N+M$ while 
disproportionation results into 2 dead chains of length $N$ and $M$.

\figitem{transfer}

Schematic of a chain transfer reaction with monomer as the transfer agent.

\figitem{hutchmwd}

MWD of poly(methyl methacrylate) produced at 40$^o$C by PLP with $t_0
= .1$sec as measured in ref. \cite{hutchinson:plp:methacrylates}.
Horizontal axis is molecular weight.  Arbitrary units on vertical
axis.

\figitem{ideal_plp}

(a) Rate of radical production as a function of time.  $\Ri(t)$ is
non-zero only during the short duration of the laser pulse the
period of which is $t_0$.  (b) Snapshot of the living chain MWD
$\phi(N,t)$ during $0<t<t_0$ in the ideal case (absence of chain transfer and
Poisson broadening). Here $\phi_0 \equiv \phi(0)$.  $\phi$ consists of
a series of $\delta$-function pulses whose amplitude follows the
envelope $\phienv(N)$.  (c) Number density of living chains as a function of
time (see eq. \eqref{fat}).  (d) Dead MWD in the ideal case.
$\phideadideal$ is discontinuous at integer multiples of $n_0$.

\figitem{dead_trans}

Contributions to the dead MWD in the presence of chain transfer.  The
upper curve is the {\em coherent} contribution.  The lower curve
is the {\em incoherent} contribution, proportional to the small
parameter $\varepsilon$.  The slope of the incoherent MWD is
discontinuous at integer multiples of $n_0$.  Values of parameters
used: $\vp = 5000 s^{-1}$, $\kc=\kd=2\,10^7 M^{-1} s^{-1}$, $\lambda =
.5 s^{-1}$, $t_0 = .1 s$, $\Ribar t_0 = 5\, 10^{-7} M$

\figitem{broad}

Snapshot of living MWD including the effects of Poisson broadening:
the $\delta$-functions of the living MWD of fig. \ref{ideal_plp}(b)
are replaced by Gaussians.

\figitem{dead_broad}

Illustration of dead MWD before and after broadening.  The solid line
is the unbroadened MWD.  The dashed line is the Poisson broadened MWD.
The dotted line is the reading on the GPC equipment, given the dashed
line as the input MWD.  We assumed uniform resolution $\sigma^2 =
5000$ (see main text).  Values of parameters used: $\vp = 5000
s^{-1}$, $\kc=\kd=5\,10^6 M^{-1} s^{-1}$, $\lambda = .1 s^{-1}$, $t_0
= .1 s$, $\Ribar t_0 = 5\, 10^{-7} M$.

                     \end{thefigures}


\pagebreak

\begin{figure}[t]

\includegraphics[width=14cm]{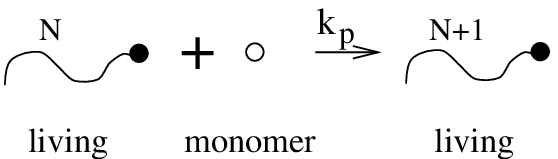}

\end{figure}

\mbox{\ }

\vfill

\addtocounter{fignumber}{1}
\mbox{\ } \hfill {\huge Fig.\@ \thefignumber} 

\pagebreak
\begin{figure}[t]

\includegraphics[width=14cm]{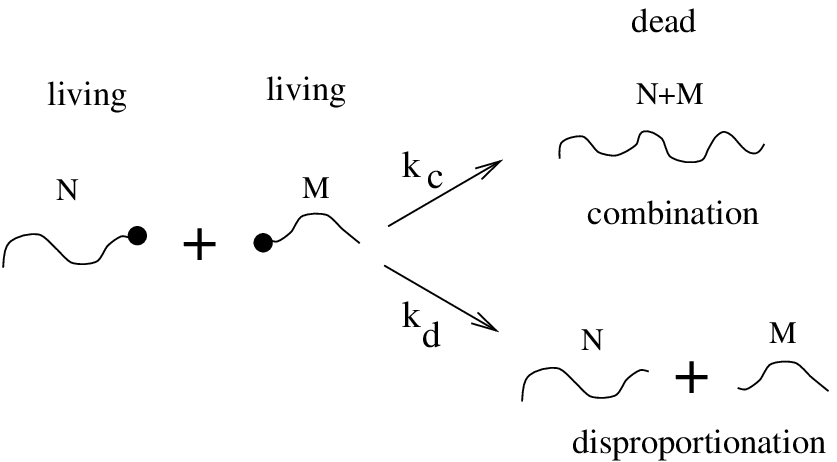}

\end{figure}

\mbox{\ }

\vfill

\addtocounter{fignumber}{1}
\mbox{\ } \hfill {\huge Fig.\@ \thefignumber} 

\pagebreak
\begin{figure}[t]

\includegraphics[width=14cm]{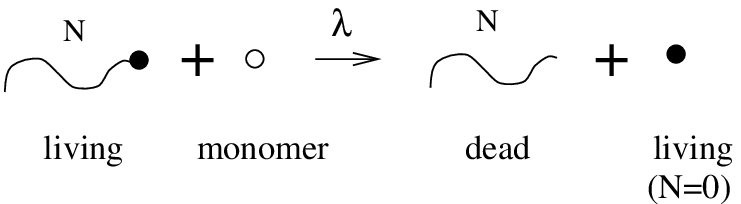}

\end{figure}

\mbox{\ }

\vfill

\addtocounter{fignumber}{1}
\mbox{\ } \hfill {\huge Fig.\@ \thefignumber} 

\pagebreak
\begin{figure}[t]

\includegraphics[width=12cm]{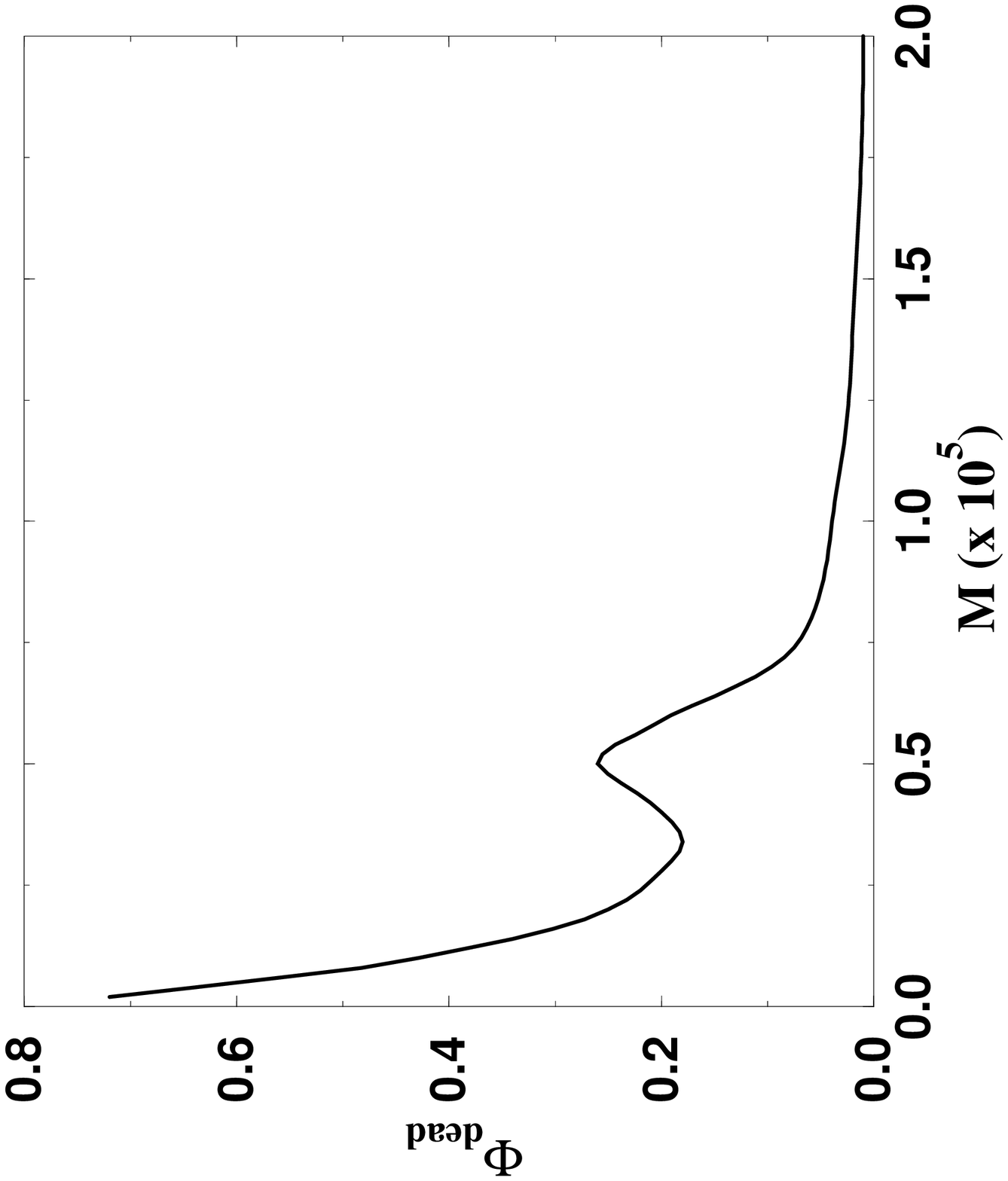}

\end{figure}

\mbox{\ }

\vfill

\addtocounter{fignumber}{1}
\mbox{\ } \hfill {\huge Fig.\@ \thefignumber} 

\pagebreak
\begin{figure}[t]

\includegraphics[width=11cm]{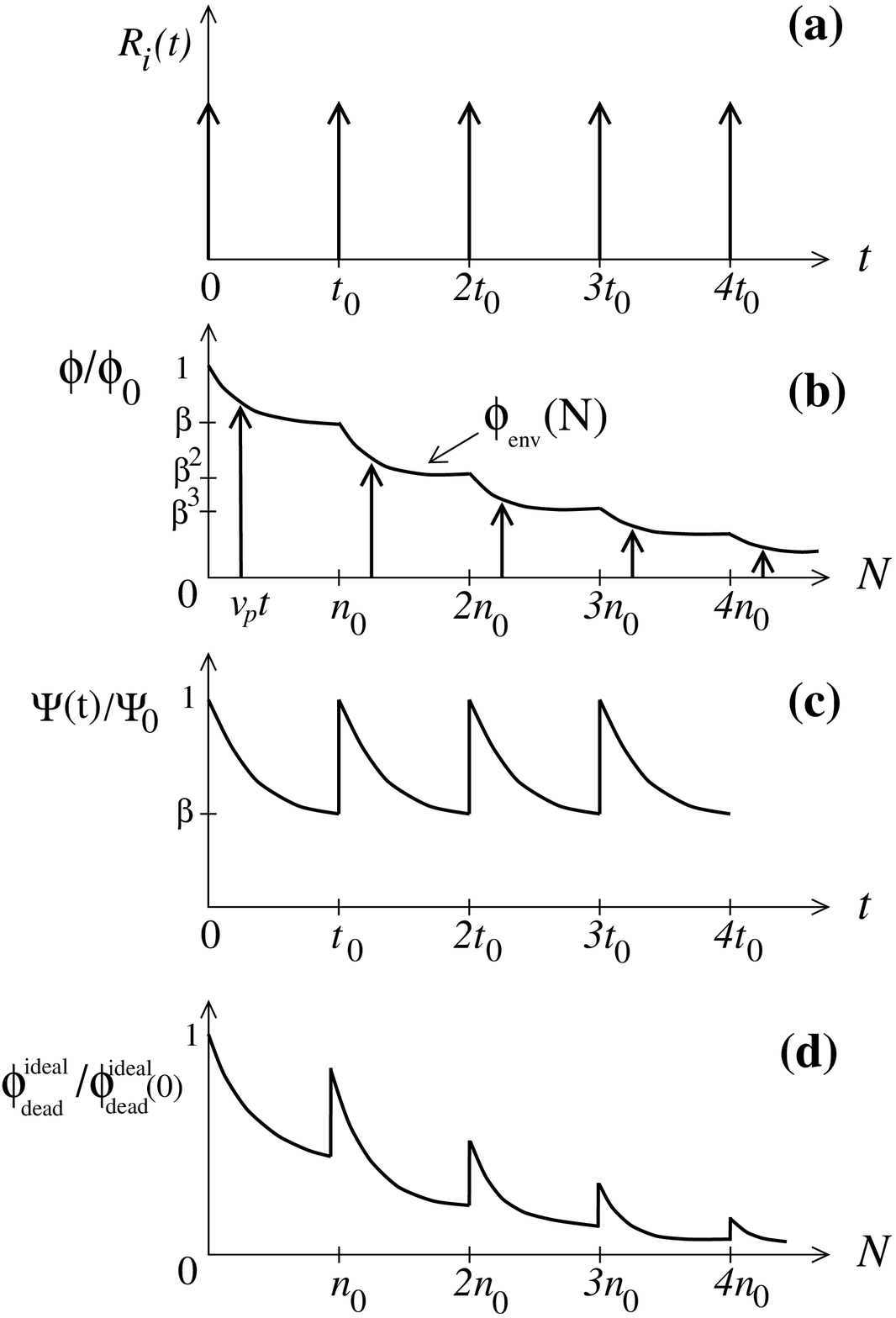}

\end{figure}

\mbox{\ }

\vfill

\addtocounter{fignumber}{1}
\mbox{\ } \hfill {\huge Fig.\@ \thefignumber} 
\pagebreak
\begin{figure}[t]

\includegraphics[width=14cm]{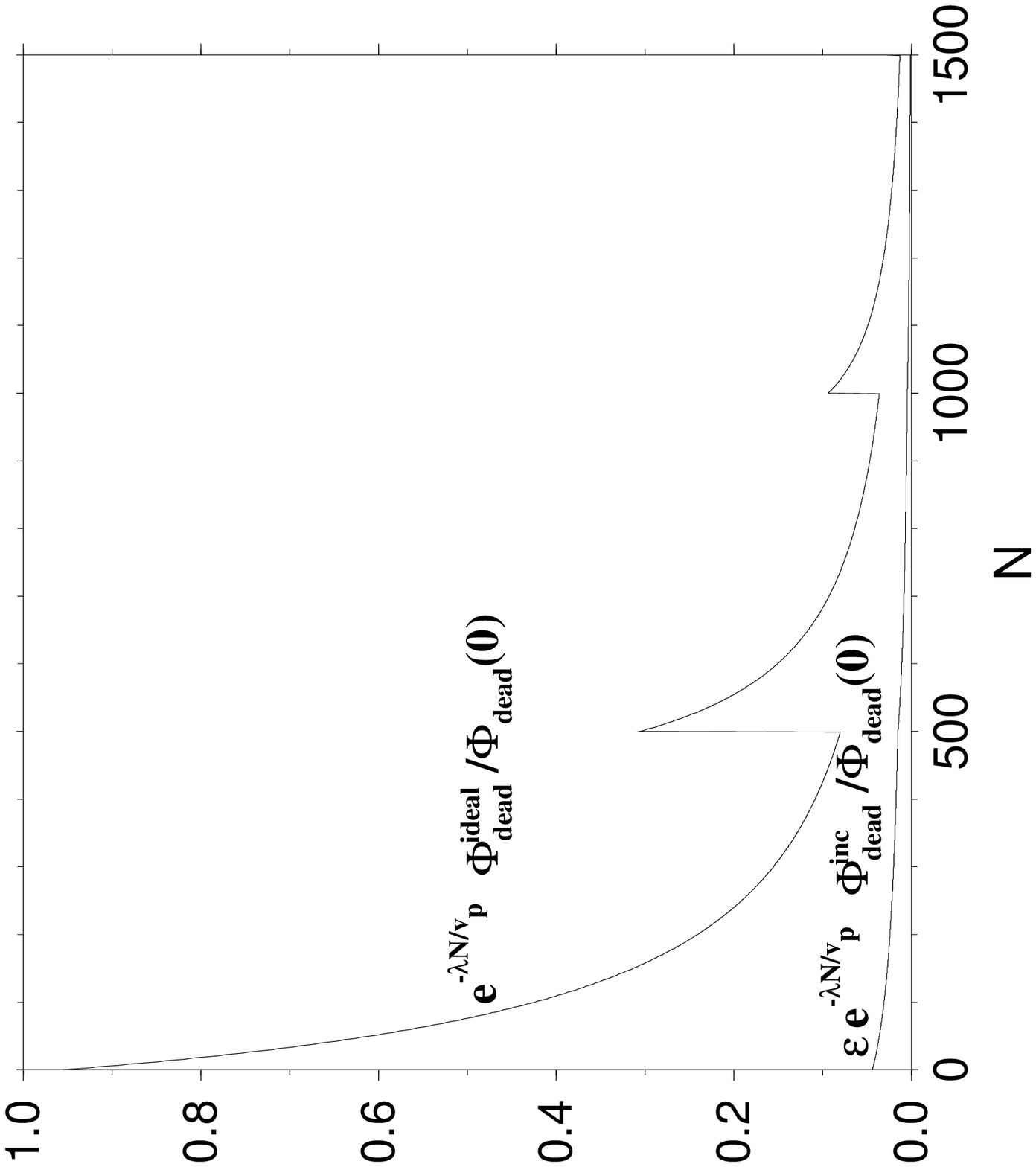}

\end{figure}

\mbox{\ }

\vfill

\addtocounter{fignumber}{1}
\mbox{\ } \hfill {\huge Fig.\@ \thefignumber} 
\pagebreak
\begin{figure}[t]

\includegraphics[width=14cm]{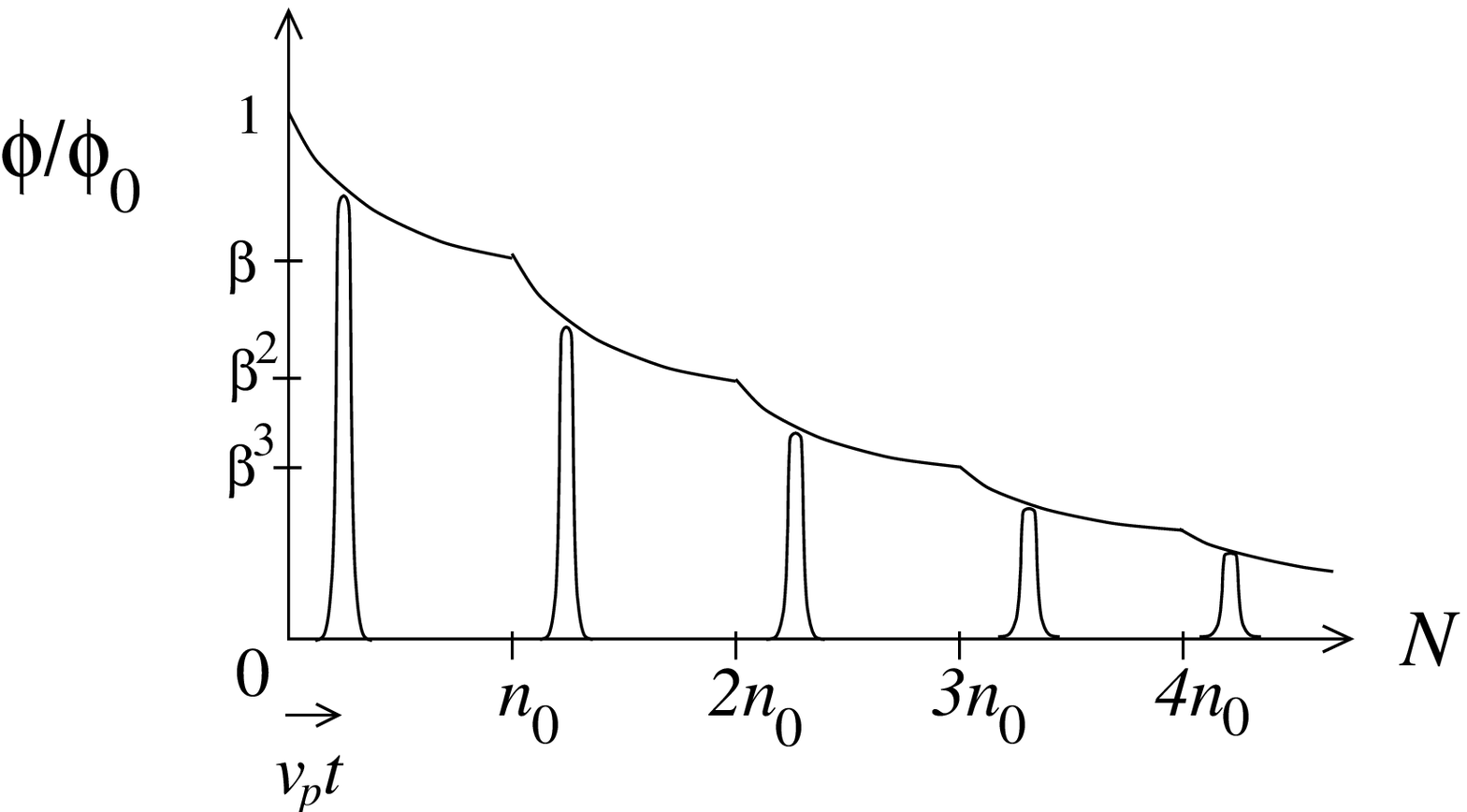}

\end{figure}

\mbox{\ }

\vfill

\addtocounter{fignumber}{1}
\mbox{\ } \hfill {\huge Fig.\@ \thefignumber} 
\pagebreak
\begin{figure}[t]

\includegraphics[width=14cm]{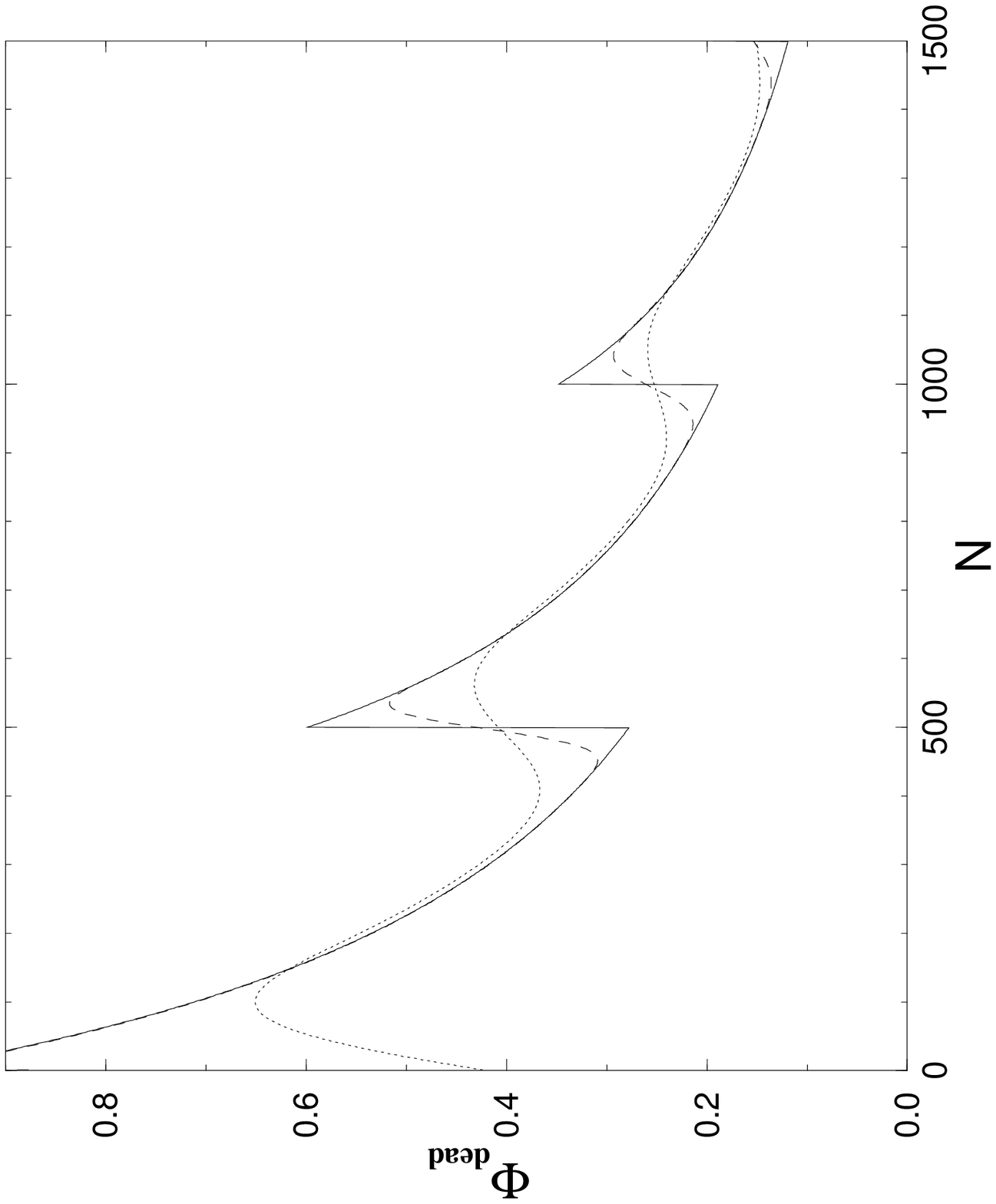}

\end{figure}

\mbox{\ }

\vfill

\addtocounter{fignumber}{1}
\mbox{\ } \hfill {\huge Fig.\@ \thefignumber}






\end{document}